# Measuring the Macroeconomic and Financial Stability of Bangladesh

Faruque Ahamed[1] and Md Ataur Rahman Chowdhury[2]

**Abstract**

This study constructs an Aggregate Financial Stability Index (AFSI) for Bangladesh to evaluate the systemic health and resilience of the country's financial system during the period 2016–2024. The index incorporates 19 macro-financial indicators across four key sectors: Real Sector, Financial and Monetary Sector, Fiscal Sector, and External Sector. Using a normalized scoring approach and equal weighting scheme, sub-indices were aggregated to form a comprehensive measure of financial stability. The findings indicate that while the Real and Fiscal sectors demonstrated modest improvements in FY2024, overall financial stability deteriorated, largely due to poor performance in the Financial and Monetary Sector and continued weakness in the External Sector. Key stress indicators include rising non-performing loans, declining capital adequacy ratios, weak capital market performance, growing external debt, and shrinking foreign exchange reserves. The study highlights the interconnectedness of macro-financial sectors and the urgent need for structural reforms, stronger regulatory oversight, and enhanced macroprudential policy coordination. The AFSI framework developed in this paper offers an early warning tool for policymakers and contributes to the literature on financial stability measurement in emerging economies.

## Introduction

Financial stability is widely regarded as a cornerstone of macroeconomic health, ensuring the efficient functioning of the financial system, safeguarding public trust in monetary and fiscal policy, and promoting sustainable development. Globally, central banks aim to preserve financial stability by maintaining a resilient financial sector and controlling systemic risks. In Bangladesh, the importance of financial stability has intensified in recent years due to growing structural vulnerabilities, increasing loan defaults, and long-standing corruption within the banking sector.

The challenges facing Bangladesh's banking system are both deep and persistent. Over the past decade, the financial sector has been marred by political interference, regulatory weaknesses, and institutional corruption. Notable financial scandals—such as the multi-billion-taka embezzlement by Prashanta Kumar Halder from several non-bank financial institutions—have exposed serious lapses in oversight and governance (The Daily Star, 2021). In addition, irregularities at banks like Padma Bank, once restructured under government support, have further eroded public confidence in the financial system (Financial Express, 2023).

---

[1] Macroeconomic Researcher, Safwa USA. email: Faruque.ahamed@safwausa.com
[2] Economic researcher, Central Bank of Bangladesh, email: ataur152@gmail,com

One of the most critical indicators of this crisis is the high volume of non-performing loans (NPLs). According to Bangladesh Bank data, the total amount of defaulted loans stood at BDT 2.11 trillion in June 2024, accounting for 12.56% of all outstanding credit (TBS News, 2024). By March 2025, this figure had surged to over BDT 4.20 trillion, representing 24.1% of total lending—within just three months, with defaulted loans rising by BDT 745.70 billion (TBS News, 2025a). Analysts estimate that without significant policy reforms, this ratio may exceed 30% soon (TBS News, 2025b). Such elevated levels of default constrain credit flow to the private sector, discourage investment, and undermine overall financial sector stability.

Bangladesh also experienced a major political turning point in 2024. Following mass student-led protests discriminatory public service recruitment quotas, the long-standing government resigned, culminating in the appointment of a transitional administration led by Nobel Laureate Muhammad Yunus (Financial Times, 2024). The new government quickly launched a series of investigations into the banking sector, uncovering alleged embezzlements of nearly USD 17 billion by individuals linked to the former regime (Reuters, 2024). These investigations revealed previously concealed systemic risks and have since led to a significant reclassification of bank loans.

Although painful, these developments present an opportunity to restore transparency and rebuild trust in financial governance. However, the interconnectedness of global financial systems means Bangladesh remains vulnerable to external shocks as well. The 2007–2009 global financial crisis demonstrated how local vulnerabilities can escalate into global disruptions through financial contagion. Despite Bangladesh's relatively limited exposure to global capital markets, volatility in trade, remittance flows, and commodity prices still pose significant risks.

This study constructs the first Aggregate Financial Stability Index (AFSI) for Bangladesh—an integrated composite measure designed to capture the health and stability of the country's financial system. By aggregating indicators from macroeconomic, banking, and capital market dimensions, the AFSI provides a robust early warning system for policymakers, regulators, and financial analysts. It serves as a practical tool to assess systemic risk, monitor the effectiveness of reform efforts, and enhance resilience against future financial shocks. In a period defined by fiscal stress, governance reform, and institutional rebuilding, such an index offers timely insights for managing stability and guiding Bangladesh toward a more transparent and sustainable financial future.

## Literature Review

Over the past decade, the concept of financial stability has garnered significant attention in both academic and policy-making circles, particularly in the context of developing economies like Bangladesh. With rising concerns over high non-performing loans, weak regulatory enforcement, and the fragility of financial institutions, researchers have increasingly sought to understand the determinants, indicators, and frameworks necessary to assess and strengthen financial stability. The literature spans a wide array of themes, including the construction of financial stability indices, stress testing, macroprudential policy tools, Islamic banking stability, capital adequacy, and the impact of governance and disclosure practices. Moreover, studies have highlighted how financial inclusion, political risk, and regulatory reforms shape the resilience of banking sectors all over the world.

Financial stability is a core responsibility of the Central Bank of Jordan (CBJ), complementing its role in maintaining monetary stability. In the wake of the 2007–2009 global financial crisis and the associated economic slowdown, Jordan's banking sector was left in a more fragile state than before. In response to these developments, Al-Rjoub (2021) constructed a comprehensive Financial Stability Index (FSI) to assess the resilience of Jordan's banking sector to economic shocks. The index aggregates fifteen officially recognized soundness indicators into three key categories: (i) capital adequacy, (ii) earnings and profitability, and (iii) liquidity. These categories were then merged into a composite index using two different weighting schemes. The study found that the FSI effectively captured the sector's response to negative shocks and shifting economic conditions. Furthermore, the index proved valuable for policymakers, offering a reliable tool to monitor systemic risk and anticipate sources of financial stress. Overall, the results indicated that the Jordanian banking system remained notably resilient in the face of economic turbulence.

Kocisova (2015) developed an Aggregate Banking Stability Index (BSI) to assess the financial stability of European Union member states, particularly focusing on the ten countries that joined the EU in 2004. The BSI was constructed as a weighted composite of various indicators, using data exclusively from commercial banks. It incorporated four key dimensions: bank performance, capital adequacy, credit risk, and liquidity risk—each assigned equal weight. Her findings revealed a notable decline in banking stability across the EU between 2005 and 2008, followed by a gradual recovery beginning in 2009.

Similarly, Karanovic and Karanovic (2015) formulated an aggregate financial stability index for the Balkan region covering the period from 1995 to 2011. Their framework included four sub-indices: a financial development index, a financial vulnerability index, a financial soundness index, and an index capturing the global economic environment. A notable contribution of their study was the incorporation of index volatility as a tool for crisis prediction.

Popovska (2014) designed a financial stability index specifically for Macedonia's banking sector by tailoring her methodology to local economic conditions. Her approach was based on the CAMELS framework, selecting only the most relevant indicators for that context. She grouped her chosen indicators into six sub-indices: capital adequacy, asset quality, management quality, profitability, liquidity, and sensitivity to interest rate and market risk, which were then combined to form the aggregate index.

## Data and Methodology

The Aggregate Financial Stability Index (AFSI) serves as a comprehensive tool for evaluating the overall stability of a country's financial system and detecting the potential buildup of systemic stress. Conceptually, it captures the degree of volatility within the financial system, recognizing that sustained deviations—either upward or downward—may signal emerging systemic risks, particularly when assessed alongside all relevant macro-financial data. The AFSI is constructed from a composite of 19 distinct indicators, which are grouped into four key sub-sectors: the Real Sector (RS), the Financial and Monetary Sector (MS), the Fiscal Sector (FS), and the External Sector (ES). By aggregating information across these diverse domains, the index provides policymakers with a multidimensional view of financial conditions and an early-warning mechanism for potential instability.

The data set used for this study comprises both bank-specific and macroeconomic variables spanning the period from 2016 to 2024. Bank-level data were collected from the annual financial statements of 23 commercial banks listed on the Dhaka Stock Exchange, all of which maintain significant market shares within the Bangladeshi financial system. These statements were accessed through the archives of the Bangladesh Securities and Exchange Commission (BSEC) library. The macroeconomic data were sourced from multiple authoritative institutions, including the Bangladesh Bureau of Statistics (BBS), Bangladesh Bank, the International Monetary Fund (IMF) Financial Statistics, and the World Bank's World Development Indicators (WDI). The final panel dataset consists of a balanced time-series cross-sectional structure, providing nine years of observed data for each of the selected banks, suitable for constructing a composite Aggregate Financial Stability Index (AFSI) and conducting empirical analysis on financial system dynamics.

At first historical data of 19 variables were collected from the above-mentioned sources. After that those numericalfigureswere converted into percentage form. Then the data set was normalized by $(x-x^-)/\sigma$. After that sub-indices and main index were calculated by applying formula mentioned below. Here it is mentionable that the weight of each indicator is equal. For example,the weight of an indicator is 1/19= .0526

The equations of AFSI and its sub-indices are mentioned below:

$AFSI_t$ = 0.15 × $RS$ + 0.15 × FS + 0.30×ES + 0.40 × MS

$RS$ = standardized value of (GDP $G_{rate}$*$W_d$ + AP*$W_d$ + QIIP * $W_d$ + I*$W_d$ + DC to $GDP_{ratio}$*$W_d$)

(Where, AP = Agricultural Production, QIIP=Quantum Index of Industrial Production, I =Inflation, DC = Domestic Credit)

$FS$ = standardized value of (FB to $GDP_{ratio}$*$W_d$ + GD to $GDP_{ratio}$ *$W_d$ + TR to $GDP_{ratio}$*$W_d$)

(Where, FB = Fiscal Balance, GD=Government Debt, TR =Tax Revenue)

$ES$ = standardized value of (ED to GDP*$W_d$ + R to ED*$W_d$ + CAB to GDP*$W_d$ + REER*$W_d$ + NIIP to GDP*$W_d$)

(Where, ED = External Debt, R = Reserve, CAB = Current Account Balance, REER=Real Effective Exchange Rate, NIIP=Net International Investment Position)

$MS$ = standardized value of (DCG*$W_d$ + PLR*$W_d$ + CRAR*$W_d$ +ROA*$W_d$ +CMR*$W_d$ + CR*$W_d$)

(Where, DCG= Domestic Credit Growth, PLR = Performing Loans Ratio, CRAR= Capital-to-Risk Weighted Assets ratio, ROA= Return on Assets, CMR= Capital Market Return, CR= Call Money Rate)

**Table 1: Summary of Data**

| Sub-sectors | Indicators | Mean (Std) |
|---|---|---|
| Real Sector | GDP Growth Rate | 0.0652 (0.0131) |
| | Agricultural Production | 0.0297 (0.0321) |
| | Quantum Index of Production | 0.0857 (0.0567) |
| | Inflation | 0.0680 (0.0178) |
| | Domestic Credit to GDP | 0.4038 (0.0177) |
| Monetary Sector | Domestic Credit Growth | 0.1307 (0.0231) |
| | Performing Loan Ratio (PLR) | 0.8986 (0.0135) |
| | Capital to Risk-weighted Asset Ratio | 0.1112 (0.0047) |
| | Return on Assets (ROA) | 0.0038 (0.0008) |
| | Capital Market Return | 0.0385 (0.2356) |
| | Call Money Rate | 0.0476 (0.0195) |
| Fiscal Sector | Fiscal balance to GDP | -0.0454 (0.0049) |
| | Govt. Debt to GDP | 0.1893 (0.0405) |
| | Tax Revenue to GDP | 0.0745 (0.0027) |
| External Sector | External Debt to GDP | 0.1944 (0.0317) |

|  | Reserve to External Debt | 0.5212 (0.1647) |
|  | Current Account Balance to GDP | -0.0124 (0.0149) |
|  | Real Effective Exchange Rate (REER) | 104.8600 (5.4134) |
|  | Net International Investment Position (NIIP) as % of GDP | -1.3347 (0.4508) |

## Empirical Analysis and Discussion

### Real Sector Index (RSI)

The real sector represents the core of economic activity, encompassing the production, distribution, and consumption of goods and services within an economy. It reflects the behavior of aggregate demand and supply, and its performance directly or indirectly influences all other sectors. Key indicators used to assess the stability and risk level of the real sector include the GDP growth rate, agricultural production, the Quantum Index of Industrial Production (QIIP), and the ratio of domestic credit to GDP.

During the review period (2016–2024), the Real Sector Index (RSI) showed a general upward trend relative to the previous period, indicating modest improvements in certain segments of the economy. However, the RSI was partially constrained by declining industrial performance and sluggish credit growth. The QIIP recorded poor growth over the last three consecutive years, and domestic credit expansion remained limited. Nonetheless, a substantial rise in agricultural production in FY2024 contributed to the upward movement of the RSI, despite its insufficient capacity to fully meet internal demand pressures.

The GDP growth rate remained relatively subdued, marking the second-lowest growth in the past decade—excluding the COVID-19-impacted year of 2020. Furthermore, inflationary pressures persisted despite global inflation declining during the same period. The domestic inflation rate reached a ten-year high, driven by multiple structural and external factors. These include: the global food crisis linked to geopolitical instability, rising import prices due to currency depreciation, increased transportation and production costs following energy price hikes, and an expansionary monetary stance involving high-speed money printing.

Additionally, the decline in industrial production has been exacerbated by high interest rates, an especially critical issue in an economy where industries are heavily dependent on bank financing. Other contributing factors include persistent supply shocks, and rising costs of raw materials, energy, and fuel. Collectively, these elements have strained real sector output and underscored its vulnerability to both domestic and international macroeconomic shocks.

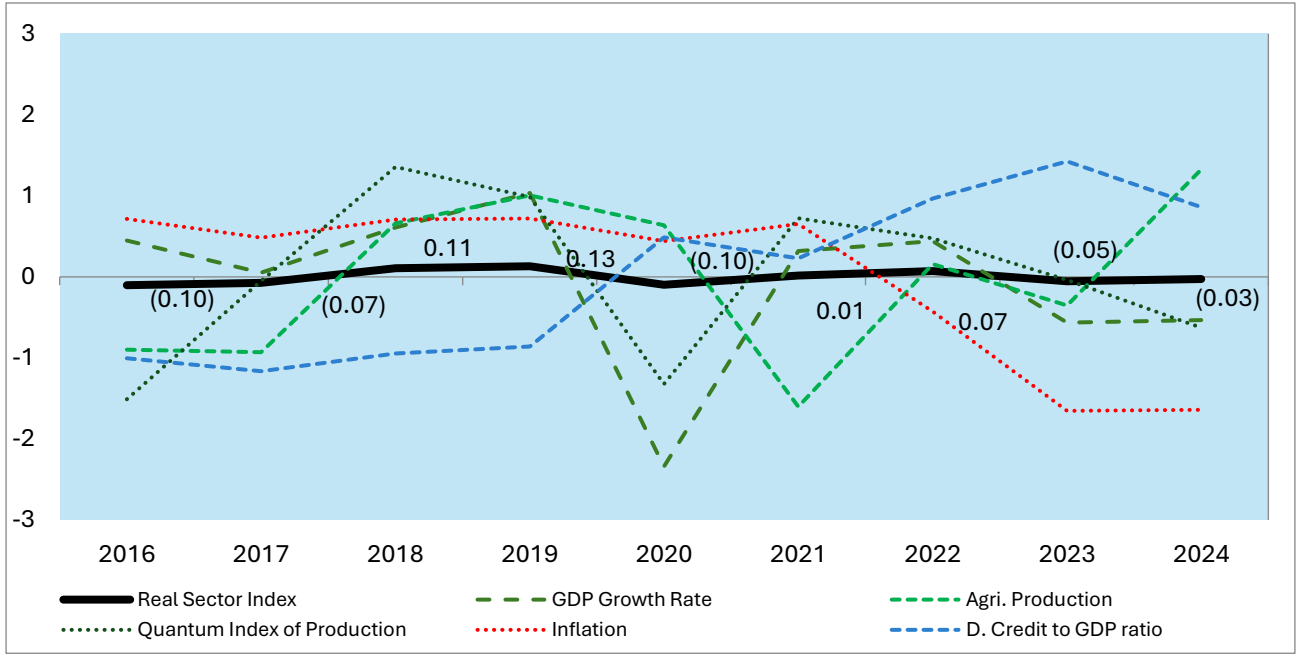

**Graph 1: Real Sector Index**

*Financial and Monetary Sector Index (MSI)*

The financial and monetary sector of Bangladesh comprises both the money market and the capital market, although it is predominantly dependent on the banking system due to the underdevelopment of the capital market. In the absence of a robust and vibrant capital market, banks remain the central financial intermediaries, providing the majority of credit and liquidity within the economy. This concentration heightens systemic risk, particularly related to maturity mismatches, which could otherwise be mitigated through more active long-term financing via the capital market. A well-functioning capital market is therefore essential for reducing risk exposure in the financial system and enhancing financial stability.

The Financial and Monetary Sector Index (MSI) used in this study is constructed from six core indicators: Domestic Credit Growth, Performing Loan Ratio, Capital to Risk-Weighted Assets Ratio (CRAR), Return on Assets (ROA), Capital Market Return, and the Call Money Rate. Across the review period (2016–2024), the MSI demonstrated a pronounced downward trend compared to the previous period, indicating increasing volatility and fragility in this sector. Although the sector temporarily showed signs of stability during the COVID-19 period primarily due to extraordinary regulatory forbearance and central bank support, this stability was short-lived.

In the post-pandemic period, multiple factors contributed to the deterioration of financial and monetary stability. A significant spike in the Call Money Rate, a sharp decline in the Performing Loan Ratio, and sustained negative returns in the capital market collectively pushed the MSI downward. Liquidity shortages within the banking system, driven by elevated inflation, rising non-performing loans (NPLs), and concerns about governance and oversight, placed significant strain on short-term lending rates. The discontinuation of temporary central bank regulatory forbearance led to a rapid and visible increase in the NPL ratio, reaching 12.56%, the highest in recent years.

Domestic credit growth also fell markedly due to a confluence of high interest rates, persistent foreign exchange shortages, and disruptions in energy supply, all of which hindered private sector investment. Meanwhile, both CRAR and ROA registered declining trends, reflecting the dual impact of reduced profitability and deteriorating asset quality. The capital market—represented primarily by the Dhaka Stock Exchange—has remained in a prolonged bearish phase, in stark contrast to many global equity markets that recovered following the COVID-19 shock and geopolitical crises such as the Russia-Ukraine war. This continued underperformance may be attributed to ongoing political and economic uncertainty, weak governance, low investor confidence, and a lack of widespread financial literacy among retail investors.

Collectively, these dynamics reflect mounting pressures within the financial and monetary sector, justifying the downward trend in the MSI and underscoring the need for institutional reforms, improved regulatory frameworks, and market deepening to enhance long-term financial system resilience.

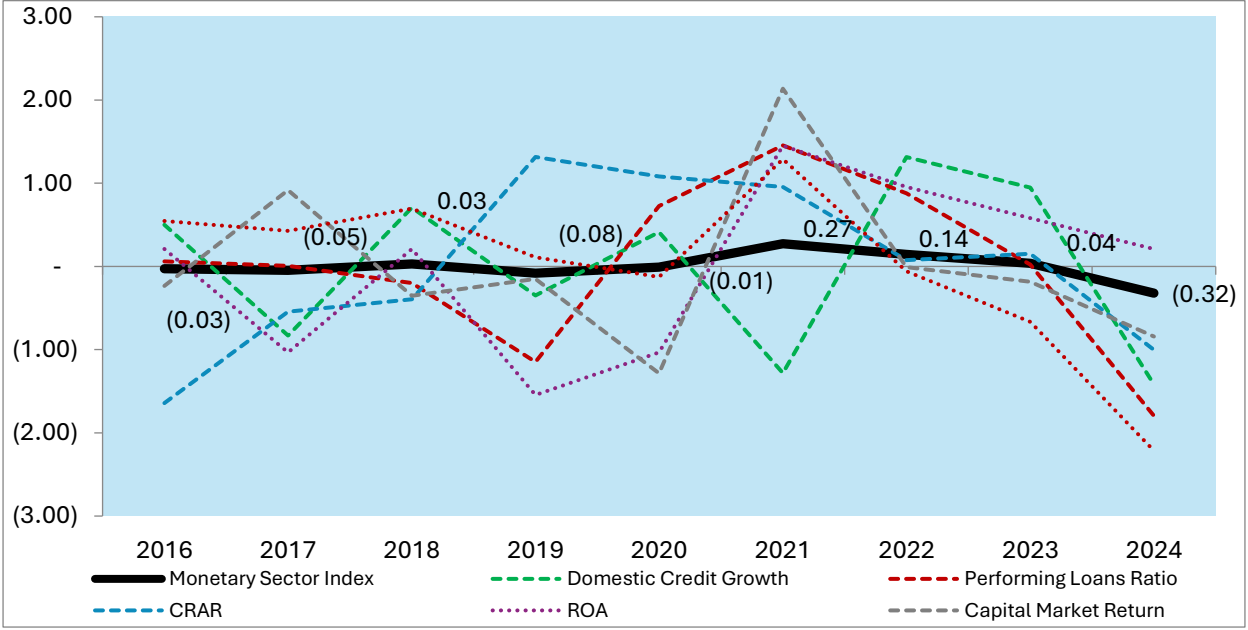

**Graph 2: Financial and Monetary Sector Index**

*Fiscal Sector Index (FSI)*

The Fiscal Sector Index (FSI) is designed to evaluate the effectiveness and sustainability of a country's fiscal policy, which is defined as the government's use of taxation and expenditure to influence macroeconomic outcomes. To capture potential vulnerabilities in this domain, the index incorporates three core indicators: Fiscal Balance to GDP, Government Debt to GDP, and Tax Revenue to GDP. These parameters collectively reflect the government's capacity to manage deficits, service its debt obligations, and mobilize domestic revenue effectively.

During the review period (2016–2024), the FSI exhibited a marginal upward trend compared to the previous period, suggesting a moderate improvement in fiscal dynamics. A notable contributor to this improvement was the increase in the Tax Revenue to GDP ratio, which provided a more sustainable revenue base and reduced dependency on debt financing. The Fiscal Balance to GDP ratio also showed signs of improvement. Although the fiscal deficit continued to expand in absolute terms, the ratio improved due to the faster nominal growth of GDP relative to the increase in fiscal deficits. This suggests that fiscal policy may have had a stimulative effect on economic expansion.

However, the rising Government Debt to GDP ratio exerted downward pressure on the FSI. Government debt experienced a significant growth rate of 25.49% during the period, outpacing the nominal GDP growth rate of 12.41%. This widening gap implies that an increasing portion of the fiscal deficit is being financed through borrowing, signaling growing fiscal vulnerability. The trend indicates a rising reliance on debt to fund expenditure commitments, which could pose risks to fiscal sustainability in the long run if left unaddressed.

Nevertheless, the upward trend in tax revenue relative to GDP provides a positive counterbalance. It reflects improvements in tax collection efficiency and possibly the expansion of the taxable base, both of which are essential for reducing fiscal imbalances over time. Overall, while the fiscal sector shows signs of stabilization, persistent debt accumulation remains a concern that requires careful policy calibration to ensure long-term fiscal resilience.

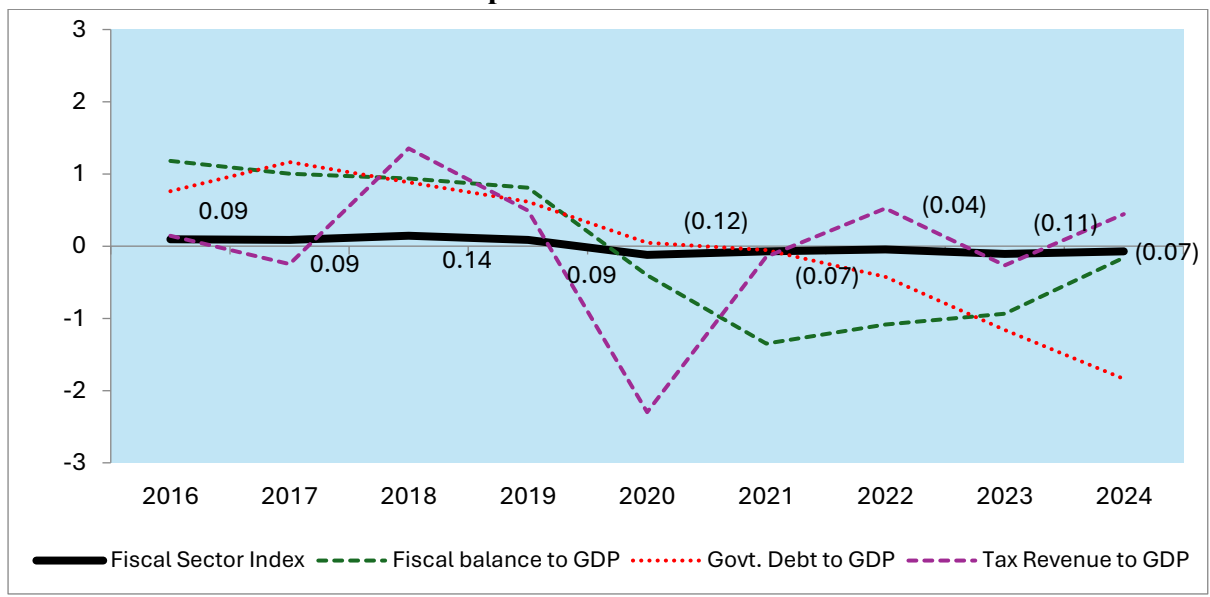

Graph 3: Fiscal Sector Index

### External Sector Index (ESI)

The external sector reflects the portion of a country's economy that engages in cross-border economic and financial transactions. In the goods and services market, this includes exports and imports, while in the financial market it encompasses capital inflows and outflows. To assess risks emerging from this sector, five key indicators are utilized in constructing the External Sector Index (ESI): External Debt to GDP, Reserve to External Debt Ratio, Current Account Balance to GDP, Real Effective Exchange Rate (REER), and the Net International Investment Position (NIIP) as a percentage of GDP.

During the review period (2016–2024), the ESI displayed a persistent downward trend relative to the preceding period, indicating growing instability in the external sector. The deterioration was primarily driven by a widening current account deficit, a declining Reserve to External Debt ratio, and a substantial rise in the External Debt to GDP ratio. A particularly concerning trend was the imbalance in trade: while imports declined by 7.58%, exports contracted by a much steeper 26.55%, leading to a significant widening of the trade gap and contributing to further deterioration in the current account balance.

At the same time, although the Real Effective Exchange Rate (REER) showed marginal improvement—reflecting a depreciation of the Bangladeshi Taka (BDT) against major foreign currencies, which slightly enhanced the competitiveness of domestic exports—this was not sufficient to offset the broader structural imbalances. The erosion of the country's foreign exchange reserves, coupled with rising external debt, led to a sharp decline in the Reserve to External Debt ratio, which fell to 19.07%. Since 2016, external debt has nearly doubled, while foreign exchange reserves have decreased by approximately 12%. This imbalance elevates perceived sovereign risk and can adversely affect the country's creditworthiness in international markets, potentially resulting in higher borrowing costs, shorter debt maturities, or restricted access to external financing.

Additionally, the Net International Investment Position (NIIP) deteriorated markedly during the review period, with its negative balance expanding faster than nominal GDP growth. This trend indicates that the country is becoming increasingly indebted to the rest of the world, undermining external sustainability. A persistently negative NIIP can weaken investor confidence and expose the economy to heightened vulnerabilities, particularly during global financial shocks.

In summary, the external sector has become increasingly fragile due to a combination of declining exports, rising external debt, and falling foreign reserves. If these trends persist, they may pose serious risks to macroeconomic and financial stability. As such, it is imperative for the central bank and policymakers to take proactive measures—such as boosting export competitiveness, managing foreign borrowing prudently, and rebuilding reserve buffers—to strengthen the external sector and restore confidence in the economy's external position.

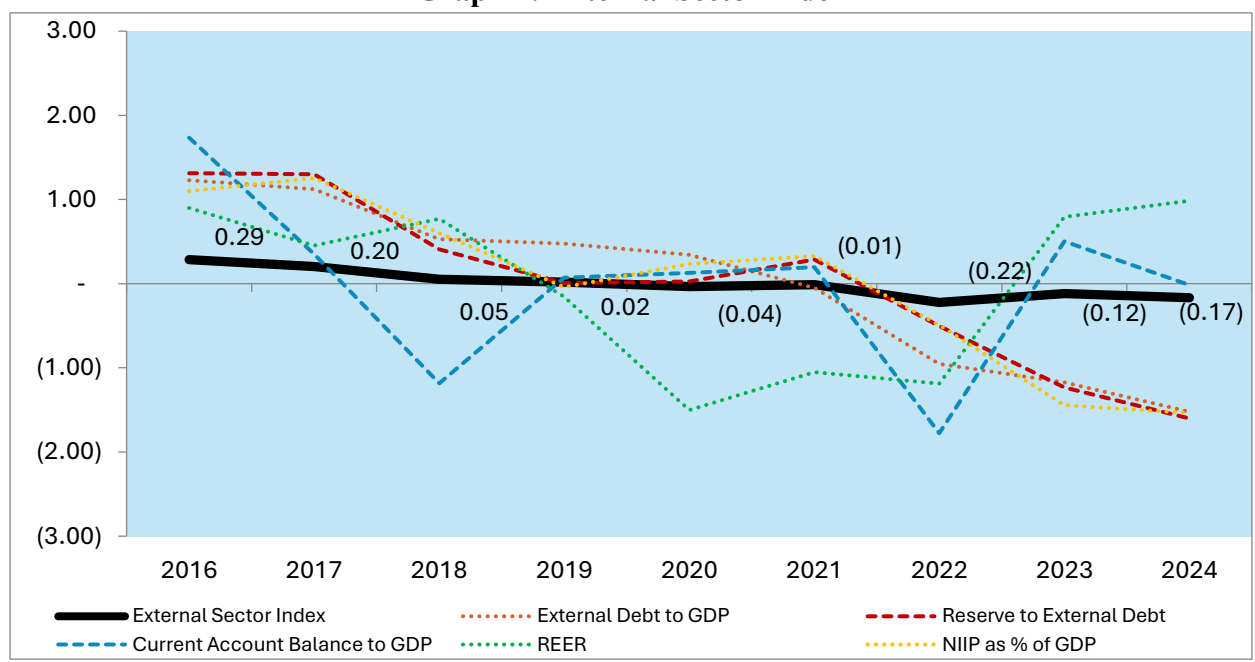

**Graph 4: External Sector Index**

*Aggregate Financial Stability Index (AFSI)*

The Aggregate Financial Stability Index (AFSI) indicates that Bangladesh's financial system experienced a sustained decline during the fiscal year 2024, largely driven by the weakened performance of the Financial and Monetary Sector. The External Sector also followed a downward trajectory, amplifying the overall decline of the Composite Financial Stability Index (CFSI). In contrast, both the Real Sector and the Fiscal Sector exhibited modest upward trends during the same period, offering some support to the broader financial system.

The sharp deterioration in the Financial and Monetary Sector Index (MSI) can be attributed to a confluence of negative indicators: a significant rise in non-performing loans (NPLs), a decline in the Capital to Risk-Weighted Assets Ratio (CRAR), poor capital market returns, and a spike in the call money rate which signal stress within the banking and financial system. Simultaneously, the External Sector Index (ESI) continued to decline, primarily due to the worsening current account balance and a notable drop in the reserve-to-external debt ratio, indicating rising external vulnerability.

On a more positive note, the Real Sector Index (RSI) improved, supported by robust agricultural output, which contributed positively to GDP growth despite persistent challenges in industrial production. The Fiscal Sector Index (FSI) also moved upward, bolstered by increased tax revenue mobilization and a narrowing fiscal balance-to-GDP ratio, even though the public debt-to-GDP ratio continued to rise. These improvements suggest a partially effective fiscal stance amidst broader macro-financial imbalances.

Notably, there exists a dynamic interdependence among the sub-indices of financial stability. The prior year's underperformance in the external sector appears to have exerted negative spillover effects on both the real and monetary sectors in FY2024. For instance, the ongoing dollar shortage and restrictive import policies led to costlier and constrained imports. This, in turn, triggered inflationary pressures and contributed to a slowdown in industrial production, affecting both the Real Sector Index and the MSI adversely.

In summary, while certain sectors demonstrated resilience, the overall financial stability of Bangladesh deteriorated during FY2024, underscoring the importance of coordinated policy responses across monetary, fiscal, and external domains to restore stability and support sustainable growth.

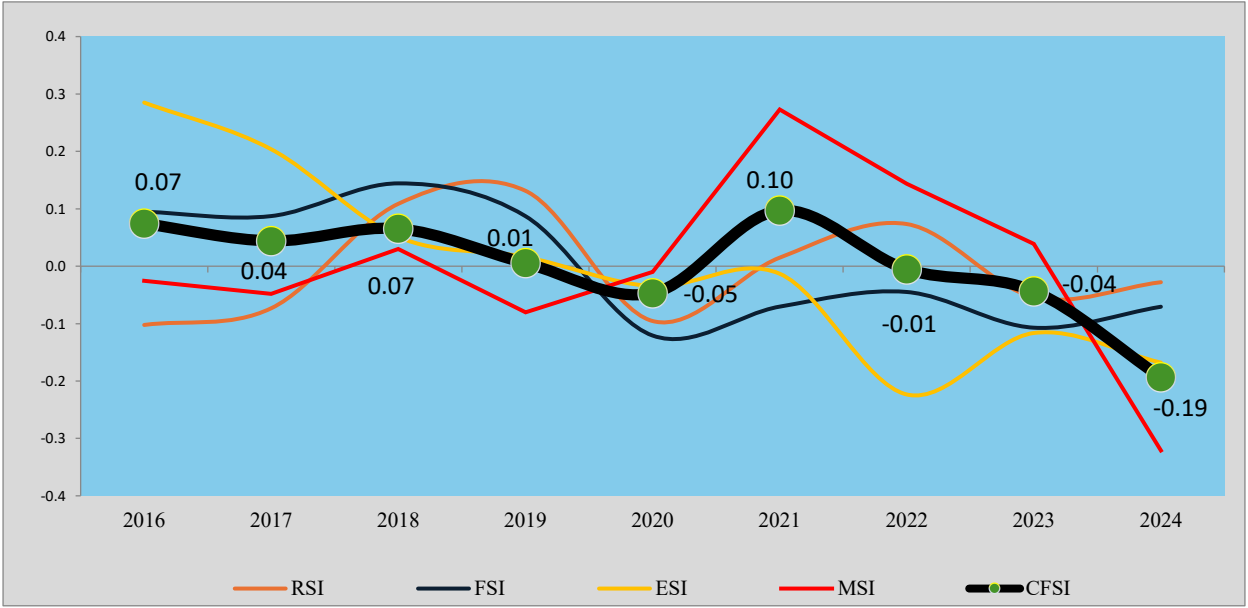

**Graph 5: Aggregate Financial Stability Index**

## Implications and Discussion

The findings from the Aggregate Financial Stability Index (AFSI) and its sub-indices—Real Sector Index (RSI), Financial and Monetary Sector Index (MSI), Fiscal Sector Index (FSI), and External Sector Index (ESI)—offer crucial insights into the current vulnerabilities and resilience within Bangladesh's financial and economic system. The overall downward trend in the AFSI during FY2024 suggests growing systemic stress, with significant implications for both macroeconomic stability and the health of the banking sector.

### *Implications for the Banking Sector*

The most concerning signals emerge from the Financial and Monetary Sector Index (MSI), which recorded a sharp decline due to multiple stress indicators. The increase in non-performing loans (NPLs), declining Capital to Risk-Weighted Assets Ratio (CRAR), falling returns on assets, and a spike in the call money rate point to liquidity constraints and deteriorating asset quality in the banking system. These developments weaken banks' lending capacity, raise the cost of funds, and erode public confidence in the financial system. As banks are the primary intermediaries in Bangladesh's financial market—due to the underdeveloped state of the capital market—instability in the banking sector has disproportionate consequences for the entire economy.

Moreover, the erosion of capital market performance and investor confidence limits the availability of long-term financing, exacerbating maturity mismatches and increasing the financial fragility of commercial banks. Inadequate credit growth amid rising interest rates and inflationary pressures also indicate that banks are increasingly risk-averse, which can slow down investment and production activities, particularly in small and medium-sized enterprises (SMEs).

### *Implications for the Broader Economy*

The economy's external and real sectors are closely linked, and the deterioration of the External Sector Index (ESI) has created significant spillover effects. The widening trade deficit, ongoing foreign exchange reserve depletion, and increased reliance on external debt have intensified balance-of-payments vulnerabilities. A depreciated currency—while boosting export competitiveness—has also raised import costs, contributing to inflation and increasing input prices for industries reliant on imported raw materials.

This external imbalance has adversely affected the Real Sector Index (RSI), particularly in terms of industrial production, which has slowed due to costly imports and supply chain disruptions. While agricultural production has provided some cushion, it is not sufficient to offset the contraction in the manufacturing sector. Persistent inflation—fueled by global food price shocks, domestic currency depreciation, and excessive monetary expansion—continues to erode household purchasing power and threatens macroeconomic stability.

On the fiscal front, the moderate improvement in the Fiscal Sector Index (FSI) reflects some positive developments such as improved tax revenue mobilization and a stable fiscal balance-to-

GDP ratio. However, rising public debt levels point to a growing reliance on deficit financing, much of which is likely sourced from banking sector borrowings. This can lead to crowding-out effects, limiting the private sector's access to credit and further slowing economic growth.

*Systemic Interlinkages and Policy Considerations*

The analysis highlights the systemic interdependence among financial sub-sectors. The weak performance of the external sector in previous years has clearly influenced the current instability in both the real and financial sectors. A shortage of foreign exchange and stringent import controls have curtailed industrial production, while banking sector liquidity has been constrained due to rising NPLs and foreign debt service requirements. Such interlinkages imply that sector-specific shocks can rapidly evolve into broader macro-financial imbalances if not addressed promptly.

To mitigate these risks, coordinated policy actions are essential. The central bank must focus on strengthening banking sector resilience through stricter loan classification standards, enhanced supervision, and recapitalization of vulnerable banks. Simultaneously, structural reforms are needed to deepen the capital market, reduce fiscal dependence on bank borrowing, and improve tax administration. On the external front, rebuilding foreign exchange reserves, promoting export diversification, and managing external debt sustainably are critical to restoring external sector health.

## Limitations of the Study

While this study provides valuable insights into the stability of Bangladesh's financial system through the construction of an Aggregate Financial Stability Index (AFSI), several limitations must be acknowledged.

The construction of the index relies on the availability and quality of secondary data collected from publicly available financial statements, central bank reports, and international databases. While efforts were made to ensure data accuracy, certain macroeconomic indicators may be subject to revisions or measurement inconsistencies, which can affect index reliability. Equal or fixed weights were applied in aggregating sub-indices and indicators. Although this approach simplifies the index construction and ensures comparability, it does not reflect the relative importance or dynamic impact of each indicator on financial stability over time. A more advanced weighting methodology—such as principal component analysis (PCA) or dynamic factor models—may offer a more precise representation but was beyond the scope of this paper.The index does not account for geopolitical risks, climate-related financial risks, or informal sector dynamics, all of which may significantly impact financial stability in a developing country context like Bangladesh.

Despite these limitations, the AFSI remains a useful tool for assessing the systemic health of the financial system and provides a foundational framework for further research, particularly in the development of early warning systems and financial risk monitoring tools in emerging economies.

## Conclusion

This study developed an Aggregate Financial Stability Index (AFSI) to assess the overall health and resilience of Bangladesh's financial system during the period 2016–2024. By integrating 19 indicators across four key sub-sectors: Real Sector, Financial and Monetary Sector, Fiscal Sector, and External Sector, the index provides a multidimensional view of systemic stability and emerging vulnerabilities. The findings reveal that while the Real and Fiscal Sectors showed modest improvements in FY2024, significant deterioration in the Financial and Monetary Sector and the External Sector led to an overall downward trend in the AFSI.

The analysis highlights several critical concerns for Bangladesh's financial architecture. The sharp increase in non-performing loans (NPLs), declining capital adequacy, weak capital market performance, and liquidity pressures have strained the banking sector, which remains the backbone of the country's financial system. Simultaneously, a widening current account deficit, shrinking foreign exchange reserves, and growing external debt have undermined external sector stability and increased sovereign risk exposure.

Despite some progress in tax revenue mobilization and agricultural output, these gains are insufficient to offset the broader macro-financial imbalances. Moreover, the strong interlinkages among sub-sectors indicate that shocks in one domain—such as the external sector—can quickly transmit to the real and financial sectors, amplifying systemic risk.

The declining AFSI underscores the urgency for comprehensive and coordinated policy action. Strengthening financial sector regulation, improving governance, diversifying exports, managing external debt prudently, and deepening the capital market are all essential steps toward enhancing Bangladesh's financial resilience. Without timely and structural reforms, prolonged financial instability could threaten sustainable economic growth and development. Therefore, the AFSI not only serves as a diagnostic tool but also as an early warning signal to guide policymakers in their efforts to ensure long-term financial and macroeconomic stability.

## References

Al-Rjoub, S. A. (2021). A financial stability index for Jordan. *Journal of Central Banking Theory and Practice, 10*(2), 157–178. https://doi.org/10.2478/jcbtp-2021-0024

Financial Express. (2023, November 3). *Padma Bank fails to return Tk 400cr deposit to state fund*. https://thefinancialexpress.com.bd/economy/bangladesh/padma-bank-fails-to-return-tk-400cr-deposit-to-state-fund

Financial Times. (2024, August 2). *Bangladesh's unlikely revolutionaries: An 84-year-old and some students*. https://www.ft.com/content/1c77dfa3-ba86-465e-bd89-1411e111237e

Karanovic, G., & Karanovic, B. (2015). Financial stability index for the Balkan region. *Economic Research-Ekonomska Istraživanja, 28*(1), 705–719. https://doi.org/10.1080/1331677X.2015.1084239

Kocisova, K. (2015). Composite index as an indicator of the financial stability of the banking sector. *Procedia Economics and Finance, 26*, 900–906. https://doi.org/10.1016/S2212-5671(15)00909-7

Popovska, J. (2014). Constructing a financial stability index for the Macedonian banking sector. *Journal of Central Banking Theory and Practice, 3*(1), 65–84. https://doi.org/10.2478/jcbtp-2014-0005

Reuters. (2024, November 17). *Around 1,500 killed in Bangladesh protests that ousted PM Hasina*. https://www.reuters.com/world/asia-pacific/around-1500-killed-bangladesh-protests-that-ousted-pm-hasina-2024-11-17/

TBS News. (2024, June 27). *Default loans climb to Tk2.11 lakh crore, cross 12.5% of total lending*. https://www.tbsnews.net/economy/banking/default-loans-climb-tk211-lakh-crore-cross-125-total-lending-1117450

TBS News. (2025a, March 29). *Bad loans soar by Tk74,570cr in 3 months to hit Tk4.2 lakh crore*. https://www.tbsnews.net/economy/banking/bad-loans-soar-tk74570cr-3-months-hit-tk42-lakh-crore-1165836

TBS News. (2025b, April 2). *Rising default loans threaten jobs, growth, and trade*. https://www.tbsnews.net/economy/rising-default-loans-threaten-jobs-growth-trade-1166861

The Daily Star. (2021, January 25). *PK Halder's embezzlement crosses Tk 3,500cr: ACC tells HC*. https://www.thedailystar.net/backpage/news/pk-halders-embezzlement-crosses-tk-3500cr-acc-tells-hc-2033997